\documentclass[twocolumn,twoside,slac_two]{revtex4}
\usepackage{graphicx}
\usepackage{fancyhdr}
\usepackage{hyperref}
\hypersetup{
    colorlinks=true,
    urlcolor=magenta,
}

\begin{document}

%Title of paper
\title{On the inverse problem for extragalactic cosmic ray nuclei with energies $10^{18}$ to $10^{20}$ eV }

% Repeat the \author .. \affiliation  etc. as needed
%
% \affiliation command applies to all authors since the last
% \affiliation command. The \affiliation command should follow the
% other information

\author{V.N.Zirakashvili, S.I.Rogovaya, V.S.Ptuskin, E.G.Klepach}
\affiliation{Pushkov Institute of Terrestrial Magnetism, Ionosphere
and Radiowave Propagation, 108840 Moscow Troitsk, Russia}

\begin{abstract}

The inverse problem of cosmic ray transport of ultra-high energy
cosmic rays is considered. The analysis of Auger data on energy
spectrum, energy dependence of mean logarithm of atomic mass number
and its variance allows definite conclusions on the shape of the
source spectrum in the frameworks of the inverse problem approach.
The discussion on regularization procedure for considered ill-posed
problem is presented.
\end{abstract}

%\maketitle must follow title, authors, abstract
\maketitle

\thispagestyle{fancy}

% body of paper here - Use proper section commands
% References should be done using the \cite, \ref, and \label commands
% Put \label in argument of \section for cross-referencing
%\section{\label{}}

\section{Introduction}
\label{sec:intro}

The origin of cosmic rays with energies $E>10^{18}$ eV is a
key problem of cosmic ray astrophysics. The observed suppression
of cosmic ray flux at energies above $\sim 5\times 10^{19}$ eV
seems confirm the presence of the GZK cutoff predicted in
\cite{Greisen66, ZatsKuzm66} although the suppression due to the
acceleration limits in cosmic ray sources can not be excluded
\cite{AllardRev, AloisoBerBlasi13}. The occurrence of the GZK
suppression and the high isotropy of the highest energy cosmic
rays are indicative of their extragalactic origin. The list of
potential sources which could give the observed cosmic ray flux
includes active galactic nuclei, gamma-ray bursts, fast spinning
newborn pulsars, interacting galaxies, large-scale structure
formation shocks and some other objects, see reviews
\cite{Olinto10,Lemoine12,BlasiRev14} and references therein.

The present knowledge about the highest energy cosmic rays was
mainly acquired from the High Resolution Fly's Eye Experiment (HiRes),
Pierre Auger Observatory (Auger), Telescope Array experiment (TA),
and from the Yakutsk complex EAS array,
see \cite{Olinto10,Troits13,Watson14}. The mass composition of
these cosmic rays remains uncertain. The interpretation of HiRes
and TA data favors predominantly proton composition at energies
$10^{18}$ to $5\times 10^{19}$ eV, whereas the Auger data indicate that
the cosmic ray composition is becoming heavier with energies
changing from predominantly proton at $10^{18}$ eV to more heavy
 composition at about $5\times
10^{19}$ eV. The mass composition interpretation of the measured
quantities depends on the assumed hadronic model of particle interactions
which is based on not well determined extrapolation of physics from lower
energies.

The energy spectrum in extragalactic sources is commonly
determined by the trail-and-error method when one makes the
calculations of the expected at the Earth cosmic ray intensity
assuming some shape of the source energy spectrum and the source
composition. The calculations follow cosmic ray propagation from
the source to the observer, e. g. \cite{Allard}. The standard
assumption is that the source spectrum is a power law on magnetic
rigidity up to some maximum rigidity.

In our previous work \cite{ptuskin15} we showed how to inverse the procedure and
calculate the source function starting from the observed at the
Earth spectrum without ad hoc assumptions about the shape of
source spectrum. Simple cases of the source composition that includes
protons and Iron nuclei were considered. The more realistic chemical
 composition including other nuclei is considered in the present work.

\section{Solution of inverse problem for a system of cosmic-ray transport equations}
\label{sec:equation}

We use the following  transport equation for cosmic ray protons and
nuclei in the expanding Universe filled with the background
electromagnetic radiation (see \cite{ASR} for detail):
\[
-\frac{\partial}{\partial\varepsilon}
\left(\varepsilon\left(\frac{H(z)}{(1+z)^{3}}+
\frac{1}{\tau(A,\varepsilon,z)}\right)F(A,\varepsilon,z) \right)
\]
\[
-H(z)(1+z)\frac{\partial}{\partial z}\left(
\frac{F(A,\varepsilon,z)}{(1+z)^{3}}\right)
+\nu(A,\varepsilon,z)F(A,\varepsilon,z)
\]
\begin{equation}
= \sum_{i=1,2...}\nu(A+i\rightarrow
A,\varepsilon,z)F(A+i,\varepsilon,z)+q(A,\varepsilon)(1+z)^{m}
\label{eq:transport}
\end{equation}
The system of eqs. (\ref{eq:transport}) for all kinds of nuclei with
different mass numbers $A$ from Iron to Hydrogen should be solved
simultaneously. The energy per nucleon $\varepsilon=E/A$ is used
here because it is approximately conserved in a process of nuclear
photodisintegration, $F(A,\varepsilon,z)$ is the corresponding
cosmic-ray distribution function, $z$ is the redshift, $q(A,\varepsilon)$ is the
density of cosmic-ray sources at the present epoch $z=0$, $m$
characterizes the source evolution (the evolution is absent for
$m=0$), $\tau(A,\varepsilon,z)$ is the characteristic time of
energy loss by the production of $e^{-}e^{+}$ pairs and pions,
$\nu(A,\varepsilon,z)$ is the frequency of nuclear
photodisintegration, the sum in the right side of eq. (\ref{eq:transport})
describes the contribution of secondary nuclei produced by the
photodisintegration of heavier nuclei,
$H(z)=H_{0}((1+z)^{3}\Omega_{m}+\Omega_{\Lambda})^{1/2}$ is the
Hubble parameter in a flat universe with the matter density
$\Omega_{m}(=0.3)$ and the $\Lambda$-term
$\Omega_{\Lambda}(=0.7)$.

A comprehensive analysis of cosmic ray propagation in the
intergalactic space was presented in \cite{Berez06}.

The numerical solution of cosmic-ray transport equations
follows the finite differences method. The variables are the
redshift $z$ and $\log(E/A)$.

Let us introduce solution $G(A,\varepsilon;A_{s},\varepsilon_{s})$
of eqs. (\ref{eq:transport}) at $z=0$ for a delta-source
$q(A,\varepsilon)=\delta_{AA_{s}}\delta(\varepsilon-\varepsilon_{s})$.
This source function describes the emission of nuclei with mass
number $A_{s}$ and energy $\varepsilon_{s}$ from cosmic ray
sources distributed over all $z$ up to some $z_{\max }$.
The general solution of eqs. (\ref{eq:transport}) at the observer location
$z=0$ can now be presented as
\begin{equation}
F(A,\varepsilon,z=0)=\sum_{A'}\int
d\varepsilon'G(A,\varepsilon;A',\varepsilon')q(A',\varepsilon').
\label{eq:green}
\end{equation}
The observed all-particle spectrum is determined by the summation
over all types of nuclei $\sum_{A}F(A,E/A,z=0)/A$ that is

\begin{equation}
N(E)=\sum_{A,A'}A^{-1}\int
d\varepsilon'G(A,E/A;A',\varepsilon')q(A',\varepsilon').
\end{equation}

We shall assume below that source spectra of nuclei can be expressed in terms of
 one function on rigidity:

\begin{equation}
q(A,\varepsilon )=k(A)Q(\varepsilon A/Z)
\end{equation}
Here $Q(\varepsilon )$ is the source proton spectrum and coefficients $k(A)$
 determine the source chemical composition.

The set of discrete values of particle energy $\varepsilon_{i}$ is
defined to solve the transport equation numerically. The grid with
constant $\triangle \varepsilon /\varepsilon$ and with $25$
energy bins per decade is used in our calculations. Eq. (3)
in the discrete form is
\begin{equation}
N_i=\sum_{j}S_{ij}Q_{j}, \
\label{eq:greennum}
\end{equation}
\[
S_{ij}=\sum_{A,A'}\frac {Z(A')k(A')}{A'A}\triangle \varepsilon_{j}
G_{ij}(A,E_i/A;A',\varepsilon _jZ(A')/A'),
\]
where the subscript indexes $i$ and $j$ denote the corresponding
energies $\varepsilon_{i}$ and $\varepsilon_{j}$.

The source term $Q_j$ can be found
from  this set of linear eqs. (5) if the observed all particle spectrum
$N(E)$ and chemical composition of the source are known. We have already considered the case
of protons and iron nuclei in the source \cite{ptuskin15}. It was found that the
solutions of equation
(5) have a physical meaning only for a limited range of proton to iron ratio. In addition the
 solution can be unstable relative small deviations of the left hand side of Eq. (5)
so that the inverse problem is ill-posed.
We shall used the following regularization procedure \cite{Tikhonov77} for this set of equations below.

Let introduce  the functional $L$
\begin{equation}
L=\sum _{i}\left( 1-\frac {1}{N_i}\sum _{j}S_{ij}Q_j\right) ^2
+\varepsilon _R\sum _j(Q_{j-1}-2Q_j+Q_{j+1})^2
\end{equation}
Here $\varepsilon _R$ is the regularization parameter. The first term in this equation is simply
 the sum of squared relative deviations from the observable spectrum $N(E)$. For $\varepsilon _R =0$ this
functional is minimized by solutions of Eqs. (5) and and its value equals to zero.

Renormalized set of equations is found from the condition $\partial L/\partial Q_j=0$:

\[
\sum _{j}S^R_{kj}Q_j=N^R_k, \ N_k^R=\sum _i\frac {1}{N_i}S_{ik}, \ S^R_{kj}=
\]
\begin{equation}
\sum_{i}\frac {1}{N^2_{i}}S_{ik}S_{ij}+
\varepsilon _R(6\delta _{kj}-4\delta _{k,j-1}-4\delta _{k.j+1}+\delta _{k,j-2}+\delta _{k.j+2}),
\label{eq:greennum}
\end{equation}

\section{Approximation of experimental data}
\label{sec:data}

To simplify calculations and damp the spread of data points in the
measured at the Earth cosmic ray spectrum, we use its analytical
approximations.

The formula
\begin{eqnarray}
J(E)\propto E^{-3.23}, E < 5\times 10^{18}\textrm{eV}; \; \nonumber \\
J(E)\propto E^{-2.63}\times [1+\exp (\log (E/10^{19.63}\textrm{eV})/0.15)]^{-1}\times \; \nonumber \\
\exp (-(E/(1.5\times 10^{20}\textrm{eV}))^{4}), E>5\times10^{18}\textrm{eV}. \;
\label{eq:analytAuger}
\end{eqnarray}
is used in our calculations to approximate the Auger data  \cite{Auger13}.
This formula is similar to the equation suggested by
the Auger team but contains $\exp (-(E/1.5\times 10^{20}\textrm{eV})^{4})$ factor of cosmic ray flux
suppression at energies $\gtrsim 1.5\times 10^{20}$ eV.

\section{Results}

The minimal value $10^{-3}-10^{-2}$ of the parameter $\varepsilon _R$ was adjusted to provide  the smooth
positive source spectrum $Q_j$. We found that this method does not work for any chemical
composition. However the range of the chemical composition is strongly extended in comparison with
 the exact solution of Eq. (5).

The results obtained for light and heavy composition of cosmic ray
non-evolutionary sources ($m=0$) are shown in Figures 1-4. The
maximum redshift $z_{\max }=3$ was used. The coefficients $k(A)$ are
given in Table \ref{tab:k}. The light composition corresponds to the
composition of
 Galactic cosmic rays.
We adjust the heavy composition to reproduce
the Auger data on energy dependence of the mean logarithm of the
atomic mass number $\langle lnA \rangle$
calculated in the EPOS-LHC model of particle interactions in the
atmosphere \cite{Auger13}.

\begin{figure}
\includegraphics[width=7.0cm]{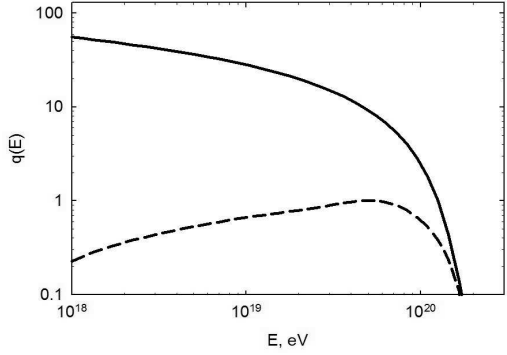}
\caption{Calculated source spectra of Iron in arbitrary units obtained for light (solid line)
and heavy (dash line) composition.}
\label{fig:source}
\end{figure}

 It is evident that our model reproduces the observed all particle spectrum and measured mean logarithm
 $\langle lnA \rangle$.

\begin{figure}
\begin{center}
\includegraphics[width=7.0cm]{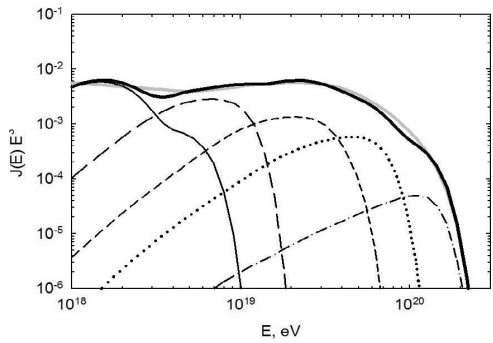}

\caption{Calculated spectra of protons (solid), He (dashed), C (short dashed), Si (dotted), Fe (dot-dashed line)
for heavy composition of  sources. The all particle spectrum (thick solid) and the analytical
approximation Auger cosmic ray spectrum
  (gray solid line) are also shown.}
\end{center}
\label{fig:calspectra}
\end{figure}

\begin{figure}
\begin{center}
\includegraphics[width=7.0cm]{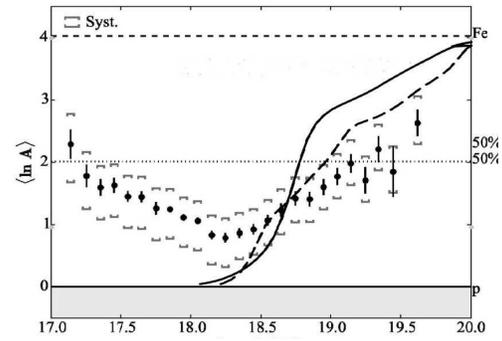}
\caption{Calculated value of $\langle ln(A) \rangle$ for light (solid line) and heavy (dashed line) composition
together with
corresponding Auger data (dots and gray regions which characterizes errors in
determination of $\langle ln(A) \rangle$ in the EPOS LHC interaction model).}
\end{center}
\label{fig:composition}
\end{figure}

\section{Discussion and Conclusion}

We showed how one can find average spectrum of extragalactic sources
from the cosmic ray spectrum observed at the Earth. This task was formulated as an inverse
problem for the system of transport eqs. (\ref{eq:transport})
that describe the propagation of ultra-high energy cosmic rays in the
expanding Universe filled with the background electromagnetic
radiation.

\begin{figure}
\begin{center}
\includegraphics[width=8.0cm]{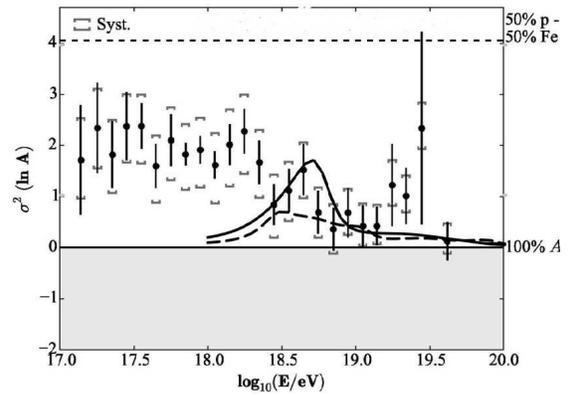}
\caption{Calculated variance of $\langle ln(A) \rangle$ for light (solid line) and heavy (dashed line) composition
together with
corresponding Auger data (dots and gray regions which characterizes errors in
determination of variance of $\langle ln(A) \rangle$ in the EPOS LHC interaction model).}
\end{center}
\label{fig:var}
\end{figure}

Mathematically, the inverse problems for transport equations
(\ref{eq:transport}) are ill-posed in the general case that
manifests itself in the instability of derived solutions. To avoid
this problem we use the regularization procedure (Eqs. 6,7)
 and perform calculations
 for a realistic chemical composition. In addition the same spectral function on the rigidity
for the source spectra of different nuclei was assumed.

We found that assumption of  heavy composition permits to explain the Auger data \cite{Auger13}.
The Auger data favor the transition from a proton source composition to the heavier one
as the energy is rising. With our heavy source composition, this case is most
closely reproduced by the calculations illustrated in figure 2. The
obtained source spectra (see figure \ref{fig:source}) resemble the results \cite{Allard08,Aloisio09}
based on the analysis of direct transport problems with a power law source spectrum.
The maximum energy of accelerated particles $(3...5)Z\times 10^{18}$ eV is
relatively low in this case that alleviates the problem of cosmic ray acceleration.
The calculated composition of cosmic rays at the Earth shown in figures 3,4
 is also in accordance with the Auger measurements.

The study of inverse transport problem is a useful tool for the investigation of ultra
high energy cosmic rays allowing the abandonment of the standard assumption of power law
source spectrum with an abrupt cutoff at some maximum magnetic rigidity as it is usually
assumed when the direct problem is considered.

\begin{table}[t]
\begin{center}
\caption{Coefficients $k(A)$ describing the chemical composition of
sources}
\begin{tabular}{|l|c|c|c|c|c|c|c|}
\hline   & H & He & C  &  O &  Mg  & Si  & Fe \\
\hline A & 1 & 4  & 12 & 16 & 24   & 28  & 56\\
\hline light & 1 & 0.2 & 0.05 & 0.05 & 0.015 & 0.04 & 0.004\\
\hline heavy & 1 & 6 & 0.65 & 0.2 & 0.1 & 0.12 & 0.015\\
\hline
\end{tabular}
\label{tab:k}
\end{center}
\end{table}

%\begin{figure}
%\includegraphics[width=65mm]{JACpic_mc.eps}
%\caption{Layout of papers.}
%\label{l2ea4-f1}
%\end{figure}

% figures should be put into the text as floats.
% Use the graphics or graphicx packages (distributed with LaTeX2e)
% and the \includegraphics macro defined in those packages.
% See the LaTeX Graphics Companion by Michel Goosens, Sebastian Rahtz,
% and Frank Mittelbach for instance.
%
% Here is an example of the general form of a figure:
% Fill in the caption in the braces of the \caption{} command. Put the label
% that you will use with \ref{} command in the braces of the \label{} command.
% Use the figure* environment if the figure should span across the
% entire page. There is no need to do explicit centering.

% \begin{figure}
% \includegraphics{}%
% \caption{\label{}}
% \end{figure}

% Surround figure environment with turnpage environment for landscape
% figure
% \begin{turnpage}
% \begin{figure}
% \includegraphics{}%
% \caption{\label{}}
% \end{figure}
% \end{turnpage}

% If you have acknowledgments, this puts in the proper section head.
\bigskip % extra skip inserted
\begin{acknowledgments}
The work was partially supported by Russian Fundamental Research Foundation grant 16-02-00255.
\end{acknowledgments}

\bigskip % extra skip inserted
% Create the reference section using BibTeX:
%\bibliography{basename of .bib file}

\end{document}